\pdfoutput=1

\documentclass[a4paper]{jpconf}
\usepackage{graphicx}

\begin{document}

\title{HS06 Benchmark for an ARM Server}

\author{Stefan Kluth}

\address{Max-Planck-Institut f\"ur Physik, F\"ohringer Ring 6, 80805 Munich, Germany}

\ead{skluth@mpp.mpg.de}

\begin{abstract}

We benchmarked an ARM cortex-A9 based server system with a four-core
CPU running at 1.1 GHz. The system used Ubuntu 12.04 as operating
system and the HEPSPEC 2006 (HS06) benchmarking suite was compiled
natively with gcc-4.4 on the system. The benchmark was run for various
settings of the relevant gcc compiler options. We did not find
significant influence from the compiler options on the benchmark
result. The final HS06 benchmark result is 10.4.

\end{abstract}

\section{Introduction}

CPUs with the ARM architecture promise to deliver more compute power per unit
of consumed energy than the curruently dominant x86 architecture CPUs.  ARM
CPUs have seen a rapid development in the recent years due to the emergence
of hand-held devices like smartphones or tablet computers where low
power consumption is essential.  The available processing capacity per core
in ARM CPUs has now advanced to a level, where their use in servers even with
large CPU power requirements has become an option and consequently ARM CPU
based server products have appeared on the market.

As a rule of thumb, in computing centres for HEP computing,
e.g.\ Tier-2 or Tier-3 centres, in Europe the investment costs for the
hardware are spent again over a period of three to four years on power
costs.  In addition, the large power consumption of big computing
clusters triggers high costs for infrastructure such a power
distribution and cooling systems.

It is therefore an interesting question, if typical tasks of todays
HEP computing could be performed with ARM based servers.  A positive
answer could lead to installations with lower power consumption and
thus also lower infrastructure costs.

We install the HEPSPEC 2006 suite and obtain benchmarks, which we then
compare with results from traditional x86 based systems.

\section{The Boston Viridis ARM server}

In this study we investigate the Boston Ltd.\ Viridis server as an
example of an ARM based server.  The Boston Viridis uses the Calxeda
EnergyCore system-on-chip (SoC) with four ARM cortex-A9 cores.  These
SoCs contain a complete system with CPU cores, 10 GBE network
interface (NIC), RAM and hard disk management.  There is up o 4~GB of RAM
for each SoC.

The SoCs are supplied on cards with four units each which are
installable in the Boston Viridis server enclosure.  In total 12 cards
with 48 SoCs may be installed.  The SoCs are supplied with power via
the cards through the enclosure with a common 300~W power supply.  The
NICs of the SoCs are connected with built-in 10 GBE switches via a
backplane in the enclosure to external LANs.

For this study we had remote administrator (root) access to a single
system housed and operated by Boston Ltd.  On this system the CPU
frequency was 1.1~GHz and it was installed with the Ubuntu 12.04 armhf
Linux OS.

\begin{figure}[htb!]
\begin{center}
\begin{tabular}{cc}
\includegraphics[width=0.4\textwidth]{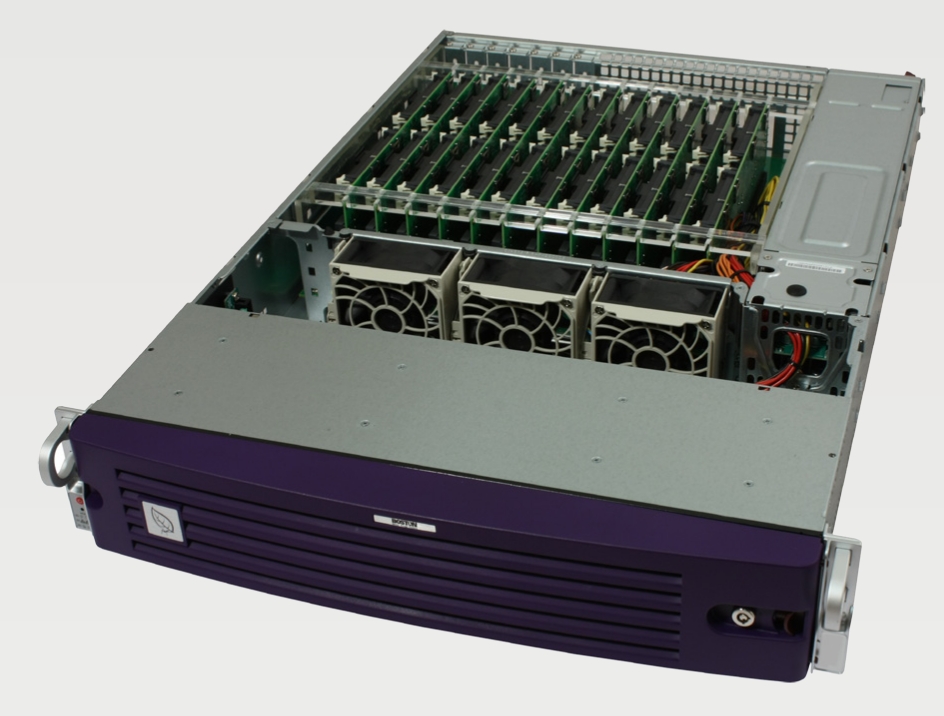} &
\includegraphics[width=0.55\textwidth]{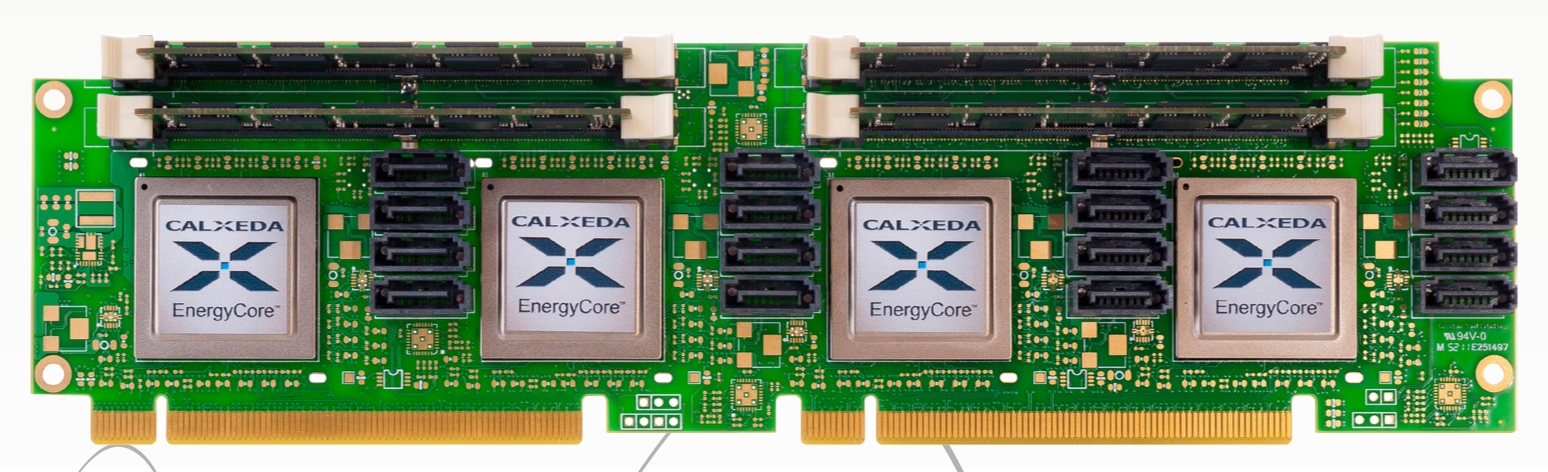} \\
\end{tabular}
\end{center}
\caption{\label{fig:viridis}(left) The Boston Viridis server enclosure. (right)
The Calexeda EnergyCard with four EnergyCore SoCs. Pictures
from~\cite{boston}. }
\end{figure}

\section{HEPSPEC 2006 benchmark installation}

The HEPSPEC 2006~\cite{hepspec2006} benchmark suite is a collection of
C++ programs from the SPEC CPU 2006 benchmark~\cite{spec2006} from
spec.org.  The benchmark runs one serial job per core and averages
individual results.  We used SPEC CPU 2006 version V1.1 together with
the spec2k6-2.23 scripts available from~\cite{hepspec2006}.

On the ARM platform the SPEC CPU 2006 toolset does not work, because
the programs are compiled for the x86 architecture.  It was thus
neccessary to compile the tools from source following the procedure
outlined in~\cite{eastlack}. 

The standard C++ compiler on Ubuntu 12.04 is gcc-4.6.  However, with
this version of gcc not all programs of the HEPSPEC 2006 suite can
be compiled successfully.  We installed gcc-4.4 from the Ubuntu 
package archives and with this version of gcc we could compile all
programs.

For the compilation of the benchmark programs we used the gcc
options \texttt{-O3 -PIC -pthread -mtune=cortex-a9} together
with \texttt{-mfpu=vfpv3} or \texttt{-mfpu=neon}.

\section{HEPSPEC 2006 results}

Table~\ref{tab:results} shows the results of our benchmark runs on the
Boston Viridis server together with results from benchmark runs on x86
systems available to us, or from other sources as indicated.  On the
x86 systems the HEPSPEC 2006 benchmark was compiled with the standard
flags for 32bit~\cite{hepspec2006} and run on 64bit Linux OS.

The main result is that the Calxeda EnergyCore SoC running at a CPU
frequency of 1.1~GHz delivers a HEPSPEC2006 value of 10.4.  In a
run with \texttt{-mfpu=neon} a value of 10.3 was obtained.
Similar results were shown at this conference by N. Neufeld.

The table also shows values for power consumption under full load
which were either obtained from manufacturers specifications in the case
of the Viridis server or from estimates based on the power envelopes given
by the CPU manufacturers and the power ratings of the power supplies.

\begin{table}[htb!]
\caption{\label{tab:results}The table shows HS06 benchmark results for
  various ARM and x86 systems together with power consumption values
  under full load and the ratios of benchmark value and power
  consumption for each system.}
\begin{center}
\begin{tabular}{lllll}
\br
System & CPU & HS06 & power [W]  & HS06/W \\
\mr
Viridis & Calxeda SoC           & 10.4 & 5 & 2.1\\
HP dc7900 & i7-2600k            & 95  & 150 & 0.63 \\
IBM HS22 & dual Xeon E5620      & 130 & 250 & 0.52 \\
IBM HS22 & dual Xeon E5645      & 179 & 250 & 0.72 \\
IBM HS23 & dual Xeon E5-2670    & 339 & 360 & 0.94 \\
Dell C6145 & dual AMD 6378~\cite{hepix} & 558 & 600 & 0.93 \\
\br
\end{tabular}
\end{center}
\end{table}

From the table one can find that the Calxeda EnergyCore platform as
used in the Boston Viridis server has a power efficiency measured in
HS06/W which is better by a factor of 2 to 4 compared with current x86
CPU based systems.  P. Szostek presented similar studies for x86 CPUs
at this conference and found similar results.  He also showed that for
the latest Intel Haswell series CPUs HS06/W values of up to 2 are
possible.

\section{Discussion and conclusions}

Based on our results it seems feasible to run HEP tasks on ARM based
servers, since sufficient processing capacity per system is available.
On the Calxeda EnergyCore platform Ubuntu and Fedora Linux are
available which makes a port of the HEP software packages possible.
N. Neufeld reported at this conference that the LHCb reconstruction
program was successfully ported to the same ARM servers as in this work
with Fedora Linux.

The investment costs for ARM based servers are still high such that
the economic case for a large scale deployment of ARM based servers is
not clear.  In addition, the Haswell generation of Intel x86 CPUs can
provide similar HS06/W ratios than the ARM server investigated here.
However, more powerful ARM CPUs supporting 64bit OS are expected on
the market soon, e.g.\ Calxeda has released the ECX-2000 SoC with ARM
cortex-A15 cores which support 64bit OS and
virtualization~\cite{calxeda}.

It still seems a good investment to study alternative platforms for
HEP computing in order to be ready when new technical and economical
possibilities arise.  Even if the outcome of such studies is negative
for the moment our ability to move to other platforms if needed sends
the right message to vendors.

\section*{Acknoledgement}

The author would like to thank Boston Ltd., UK, for providing the
remote access to a server as well as technical help for our tests.

\end{document}